\documentclass[twocolumn,secnumarabic,amssymb, nobibnotes, aps, prd]{revtex4-2}

\usepackage{graphicx, bm}
\begin{document}

\title{Genetic-Multi-initial Generalized VQE: Advanced VQE method using Genetic Algorithms then Local Search
}%

\author{Hikaru Wakaura}%
\email[Quantscape: ]{
hikaruwakaura@gmail.com}
\affiliation{QuantScape Inc. QuantScape Inc., 4-11-18, Manshon-Shimizudai, Meguro, Tokyo, 153-0064, Japan}

\author{Takao Tomono}

\affiliation{ Digital Innovation Div. TOPPAN Inc. , 1-5-1, Taito, Taito, Tokyo, 110-8560, Japan}
\email[TOPPAN: ]{takao.tomono@ieee.org}
\date{September 2021}%

\begin{abstract}

Variational-Quantum-Eigensolver(VQE) method has been known as the method of chemical calculation using quantum computers and classical computers. This method also can derive the energy levels of excited states by Variational-Quantum-Deflation(VQD) method. Although, parameter landscape of excited state have many local minimums that the results are tend to be trapped by them. Therefore, we apply Genetic Algorithms then Local Search(GA then LS) as the classical optimizer of VQE method. We performed the calculation of ground and excited states and their energies on hydrogen molecule by modified GA then LS.Here we uses Powell, Broyden-Fletcher-Goldferb-Shanno, Nelder-Mead, and Newton's method as an optimizer of LS. We obtained the result that Newton's method can derive ground and excited states and their energies in higher accuracy than others. we are predicting that newton method is more effective for  speedup and be more accurate.

\end{abstract}

\maketitle
\tableofcontents

\section{Introduction}\label{1}
Today, there are various types of Variational-Quantum-Eigensolver(VQE) methods in the world. For example, Subspace-Search VQE (SSVQE) \cite{PhysRevResearch.1.033062} can calculate the multiple energy levels at once and Multiscale-Contracted VQE (MCVQE) \cite{2019PhRvL.122w0401P} calculates the ground and single electron excited states by diagonalizing Configuration Interaction State (CIS) Hamiltonian. Adaptive VQE \cite{2019NatCo..10.3007G} and Deep-VQE \cite{2020arXiv200710917F} are also proposed. The essential procedure of quantum circuit learning has taken advantage of VQE \cite{PhysRevA.98.032309}. There is also the method that takes advantage of Genetic Algorithms(GA)\cite{Kobayashi2009}, called Evolutional-VQE(E-VQE)\cite{2019arXiv191009694R}. 
In parallel to the development, quantum hardware has been improved concerning both the number of qubits and quantum volume. The fidelity of qubits skyrocketed last year. Both Honeywell \cite{2020PhRvR...2a3317B} and Ion-Q \cite{IonQ2020} updated the record of quantum volume twice and the record of this is four million achieved by the quantum computer of Ion-Q. Recently, the Institute of Science in China has achieved quantum supremacy by photonic quantum computer \cite{Zhong1460}.

An advanced algorithm of E-VQE that uses GA then Local Search(GA then LS)\cite{Harada2006} has been demonstrated. This is Multi-objective Generalized VQE(MoG-VQE)\cite{2020arXiv200704424C}. This algorithm optimizes the clusters by GA and optimizes their variables. 
This algorithm takes much time compared to other VQE methods because two kinds of optimization must be performed for the optimization of both clusters and variables. Therefore, we propose a new VQE method using GA then LS, called Genetic Multi-Initial Generalized VQE(GMIG-VQE). This algorithm optimizes the variational parameters by GA  and chosen some parameter sets that are optimized by the local minimizer. As the result of the VQE method improve by applying the result of the MP2 method as the initial value, preparing close parameter sets to a global minimum is promised to calculate the eigenenergies of energy levels more accurately than the conventional VQE method. Time for calculation will be shorter than MoG-VQE because local search(LS) is used for only chosen parameter sets. Here, we evaluated sevral methods as LS.

The following organizations of this paper are as follows. Chapter \ref{2} is describing the detail of our method. Chapter \ref{3} indicates the result of our calculation on hydrogen molecules using GMIG-VQE. Chapter \ref{4} is the conclusion of our works.

\section{Method}\label{2}

In this section, we describe the method of GMIG-VQE. This algorithm has the same procedure as the VQE method except for the process of optimization. The ordinary VQE is a hybrid method that calculates trial energy on quantum computers for variables and optimizes the variables to find the minimum of trial energy. The equation of the trial energy is as follows,

\begin{equation}
E_i(\bm{\theta^i})=\langle \Phi_{ini} \mid U^\dagger(\bm{\theta^i})HU(\bm{\theta^i})\mid \Phi_{ini} \rangle\label{E},
\end{equation}

where $\mid \Phi_{ini} \rangle$ is initial state, $U(\bm{\theta^i})$ is the operator to make the given superposition state that includes the trotterized Hamiltonian and cluster terms correspond to  $\bm{\theta^i}$, that is the variable vector of $i^{th}$ state, respectively. $U(\bm{\theta^i})$ . The depth of Hamiltonian and cluster terms are set to $n$, which in the following discussions is 2. It is expressed as $U(\bm{\theta^i})=\prod_l \prod_k exp(-i\theta_l^i c_k^lP_k^l)$ by the index of the variable $l$ and the index of terms belongs to variable $l$ transformed into Pauli words $k$ \cite{doi:10.1021/acs.jctc.8b00450} \cite{2017arXiv171007629M}. We use Unitary Coupled Cluster(UCC) \cite{2018PhRvA..98b2322B} ansatz to calculate the derivatives. The function to be minimized in actual VQE method is a function that contains constraint terms \cite{doi:10.1021/acs.jctc.8b00943} $E^{const}_j(\bm{\theta^i})$ and deflation terms of Variational Quantum Deflation(VQD) \cite{2018arXiv180508138H} $E^{def}_j(\bm{\theta^i})$.The method to calculate the product between two states is SWAP-test \cite{2013PhRvA..87e2330G}, which is used to calculate the excited states. They are for calculation of excited states. The evaluation function of $i^{th}$ state is,

\begin{equation}
F_i(\bm{\theta^i})=E_j(\bm{\theta^i})+E^{const}_j(\bm{\theta^i})+E^{def}_j(\bm{\theta^i})\label{F}.
\end{equation}

The constraint and deflation terms are zero at a global minimum of the evaluation function. The process of optimization is GA then LS. 
It optimizes the individuals of GA and optimizes parameter sets coincidently. We use real-coded GA for GA that optimizes the parameter sets. rcGA is the one of GA that optimizes the continuum variables. We use the Real-coded Ensemble Crossover(REX) method as a crossover method.

It is expressed as, 

\begin{equation}
\bm{\theta}^{(g+1,n)}=\bm{\theta}^{(g)}+\sum_j^{N_p}\xi_j(\bm{\theta}^{(g,j)}-\bm{\theta}^{(g)}.
\end{equation}

Then, $\bar{\bm{\theta}}^{(g)}$ indicates the average of parent sets of $g$ generation and $\bm{\theta}^{(g, n)}$ indicates the $n^{th}$ children set of $g^{th}$ generation and $N_p$ indicates the number of parent sets, respectively. JGG is a generation alternation method, which is executed as follows. First, for each parameter $N$, the first generation $10N$ individuals are generated randomly. The initial values are generated according to the following equation.

\begin{equation}
\bm{\theta}_j^{(0, m)} = (\bm{\theta}_j^{UB}-\bm{\theta}_j^{LB})f_{ini} + \bm{\theta}_j^{LB}
\end{equation}

where $\bm{\theta}_j^{UB},\bm{\theta}_j^{LB}$ are the upper and lower limits of the $j$th parameter, respectively.
Also, $f_{ini}$ is the initialization function.
After the second generation, $N_p$ parents are randomly selected from the population, and $N_c$ children are generated from them.
Family set is generated after that.  Closest $N_p$ parameter sets to the aimed value of evaluation function $F_i(\bm{\theta}^{(g,j)})$ are chosen and rereased into the individuals. Others are trashed.
This is one generation so far.
Thereafter, the generation is updated iteratively.
Originally, the mean gap should be the convergence condition. However, the target evaluation value of the evaluation function is set to negative infinity, the standard deviation of each variable in $F_i(\bm{\theta}^{(g, j)})$ is set as the convergence condition. The convergence condition is

\begin{eqnarray}
\sigma_k &=& \sum_{j}^{10N}(\bm{\theta}^{(g, j)}_k - \bar{\bm{\theta}^{(g)}_k})^2/10N \\
10^{-16} &>& Max(\sigma_k/(\bm{\theta}^{UB}_k - \bm{\theta}^{LB}_k))\label{conv}.
\end{eqnarray}

Here, $\sigma_k$ is the standard deviation in the $k$th variable. In this case, the process of creating children and calculating their evaluation functions can be done in parallel, as long as the number of qubits and computational resources of the quantum computer allows, by assigning the calculation of the evaluation function for each child to an individual quantum computer as a job, which is how it is done in this paper. Combining this with a local optimization algorithm, the GA then LS was devised to improve the accuracy of optimization in the parameter space with UV structure \cite{2010D-EC0904}.
This GA then LS is a method that combines the genetic algorithm and other optimization methods with local optimization in the genetic algorithm to achieve a globally optimal solution while performing local optimization of parameters.
However, since it is too time-consuming to perform local optimization for each generation, thus, we first perform the genetic algorithm and LS after that, choosing the best 10 parameters with the smallest difference between the evaluation function and the target value as initial parameter sets of LS.

The initial function of the GA is the following function, which is a mixture of the beta distribution function and the uniform distribution in the ratio of 1000:1.

\begin{equation}
f_{ini}=(B(0.99, 0.99) + 0.001 Rand)\label{ini}
\end{equation}

Here, $B(0.99,0.99)$ and $Rand$ are the beta distribution function with $\alpha=\beta=0.99$ and the uniform distribution function with a range from 0 to 1, respectively.
Here, the initial values can exceed the defined range of the variables, but this does not pose a problem since it does not affect LS, although it does affect the genetic algorithm, and the global optimal solution can be calculated.
The main flow of GMIG-VQE is as follows.

\begin{enumerate}

\item
Create a set of initial parameters for 10N individuals according to the formula \ref{ini}. Calculate the evaluation function F for all of them.

\item
Repeat the calculation of the alternation of generations by a fixed number of crossings and generation of offspring until the standard deviation $\sigma_k$ of each variable in the evaluation function $F_i(\bm{\theta}^{(g, k)})$ satisfies the convergence condition \ref{conv} or until a specified number of alternations have been passed.

\item
Perform LS based on the parameters of the 10 individuals with the smallest value of the evaluation function $F_i(\bm{\theta}^{(g, k)})$. Perform the same calculation as the VQE method using the classical optimization method chosen in advance. Here $\bm{\theta}^{(g, k)}$ is the $k^{th}$ individual among all individuals.

\item
From the results obtained, the solution is the one with the smallest selector function, $log((F_j(\bm{\theta}^{(g,k)})-E_j)/E_j),k\in\{1,2,\cdots,10\}$.

\end{enumerate}

\begin{figure*}[h]
\includegraphics[scale=0.25]{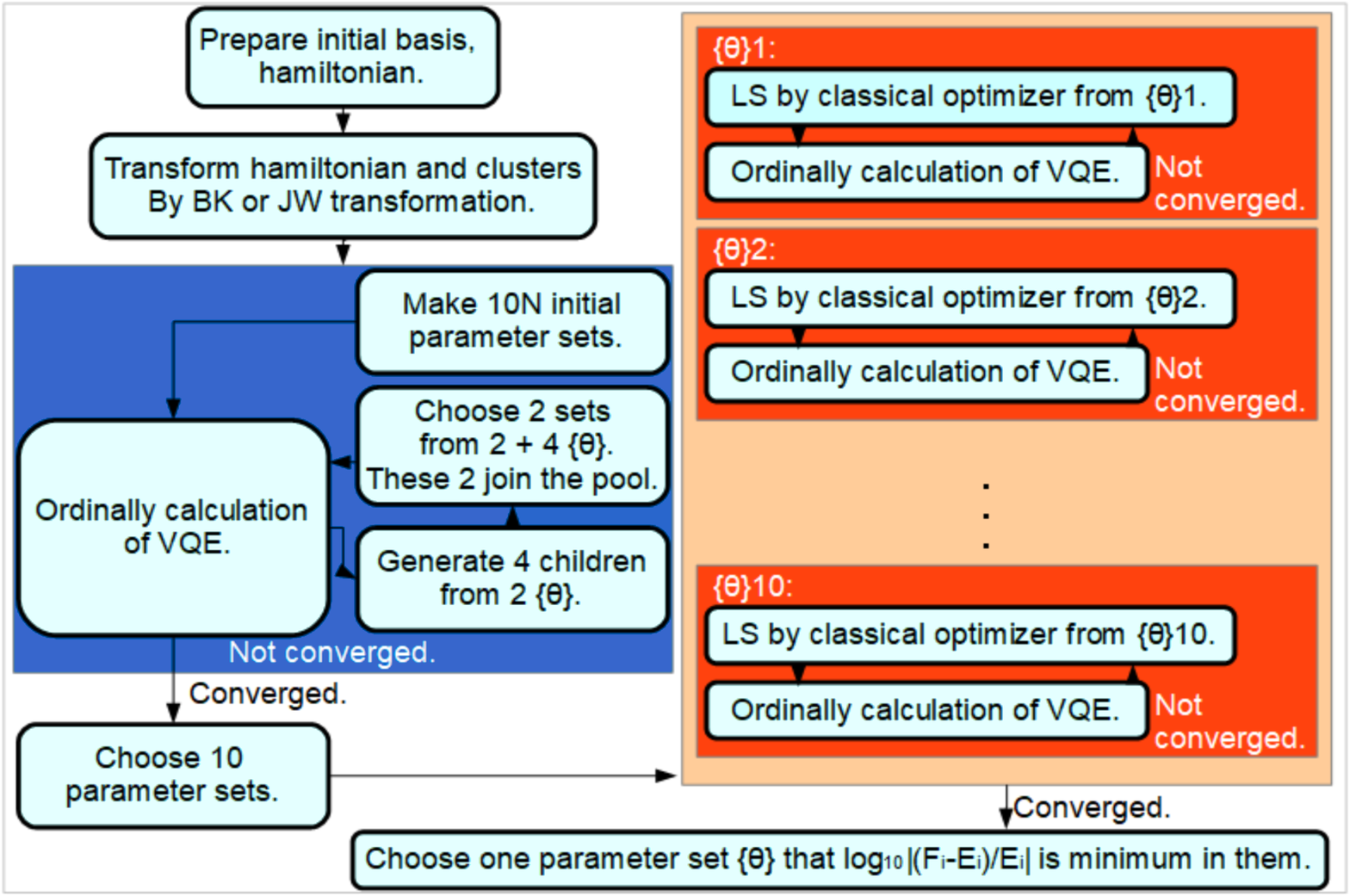}
\newline
\newline

\caption{The flowchart of GMIG-VQE that 10 parameter sets are chosen for LS. $\{\theta\}j$ indicates the $j$th parameter set in chosen 10 sets. Blue rectangle area indicates that classical optimizer of this area is GA, and orenge rectangle area indicates that classical optimizer of this area is LS, respectively.}\label{gmigvqe}
\end{figure*}

The detailed flow is shown in Fig. \ref{gmigvqe}.
This time, only the ground state is optimized for all variables in the entire process, and for the excited state, the Hamiltonian variables in the GA are the optimized values in the ground state. The Hamiltonian and the basis are STO-3G, and the depths of the cluster and the Hamiltonian are both 2. All results of quantum calculations are in the form of the state vectors(number of shots are infinity).
The actual calculations are performed using openfermion\cite{openfermion} and blueqat SDK\cite{blueqat}, which is a quantum computer simulator, and for the genetic algorithm, we used our program modified from vcopt\cite{vcopt}.

Then, $N_p$ is 2, $N_c$ is 4 and the number of the population is $10N$ for the number of dimensions of the parameter.
It is because as the $N_c$ increases, more time for GA is required. The multi-valley structure of parameter landscape is also the reason. The parameter $\xi_j$ is the random variable that average is 0 and standard deviation is $0.9/\sqrt{N_p}$. This variable is to save the distribution of population. The number of candidates in the LS is 1/5 for the hydrogen molecule since two identical values appear for each molecule.
If more than 1/5 of the total number of individuals are considered as candidates, the probability of including individuals that do not fall into the local solution will be sufficiently high.

\section{Result}\label{3}
In this section, we denote the numerical result of the calculation on energy levels of hydrogen molecules using GMIG-VQE for using Powell, Conjugate Gradient, Broyden-Fischer-Goldfarb-Shanno, and Newton's methods as LS. 

Firstly, we show the result of ordinary VQE with the VQD method as shown in Table. \ref{avea}. The number of iteration of each method are 2000 for Powell and Nelder-Mead methods, 50 for the BFGS method, and 1000 for the Newton's method, respectively. We show the average ordinary log of the difference between the calculated energies and exact value calculated by STO-3G classically. We call this log error. The log errors are shown as the result of the calculation on the ground, triplet, singlet, and doubly excited states of hydrogen molecules. The diatomic bond length is from 0.1$(\AA)$ to 2.5$(\AA)$ in 0.1 pitch. The VQE with VQD method is performed using (a)Powell, (b)Broyden-Fletcher-Goldfarb-Shanno (BFGS), (c)Nelder-Mead and (d)Newton's methods. (b)BFGS method has the smallest average of log error for all states. (d)Newton's method has the second smallest average of log error for singlet and doubly excited states. BFGS method is a quasi-Newtonian method, thus, Newtonian and quasi-Newtonian methods are suitable for VQE with VQD due to calculation on hessian or quasi-Hessian. (a) and (c) have large accuracy because these two optimizing methods update the evaluation functions only smaller.

\begin{table}[h]
\caption{
The average log error of ground, triplet, singlet, and doubly excited state on hydrogen molecule for diatomic bond length r in case (a), (b), (c), and (d), respectively. These calculation were performed by ordinary VQE.
}\label{avea}
\begin{tabular}{c|c|c|c|c} \hline \hline
method&Ground&Triplet&Singlet&Doubly \\ \hline
Powell&-10.6579&-3.7597&-1.6531&-1.1804 \\
BFGS&-11.1289&-10.739&-2.1491&-1.9588 \\
Nelder-Mead&-10.6392&-5.445&-1.5923&-1.1956 \\
Newton&-10.6269&-10.181&-2.0473&-1.655 \\ \hline

\end{tabular}
\end{table}

\begin{figure*}[h]
\includegraphics[scale=0.45]{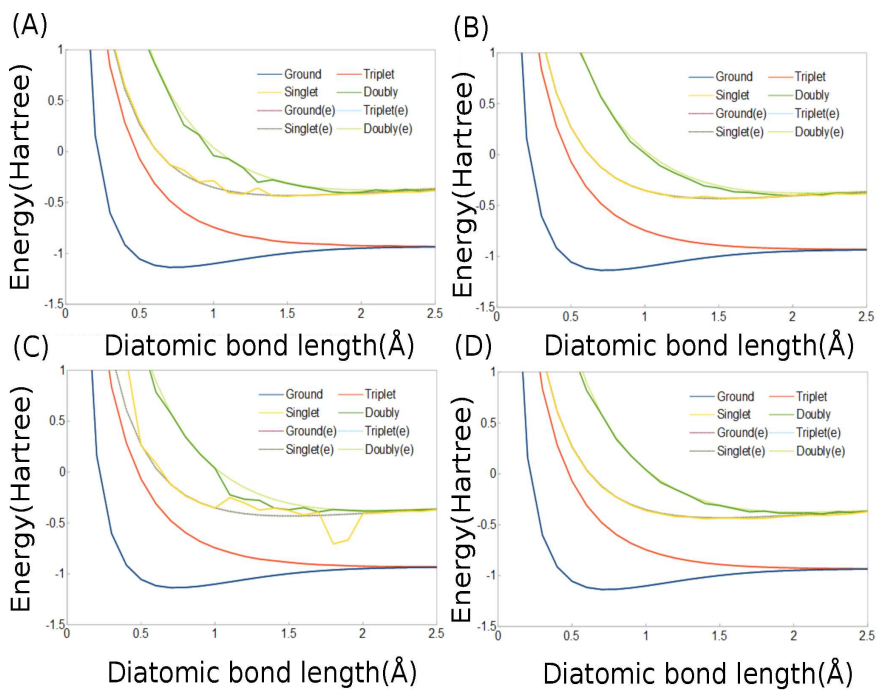}
\newline

\caption{The diatomic bond length v.s. the energy levels of ground, triplet, singlet, and doubly excited states on hydrogen molecule, respectively. Local search method is (A)Powell, (B)BFGS, (C)Nelder-Mead, and (D)Newton's methods. The lines that have (e) in their suffix are exact values calculated by the Full-CI method.}\label{gene}
\end{figure*}

\begin{figure*}[h]
\includegraphics[scale=0.45]{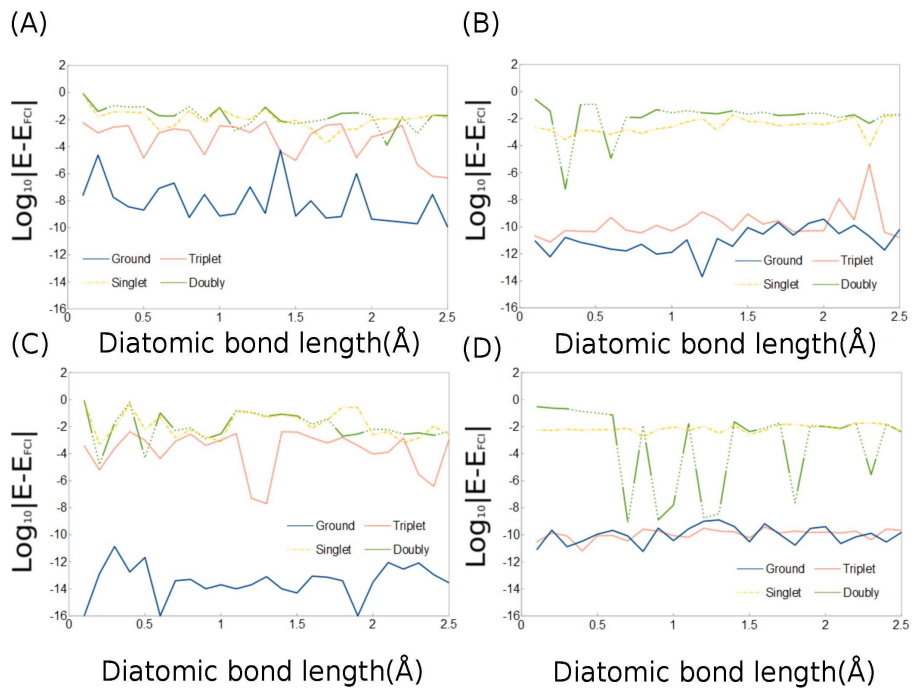}
\newline

\caption{The diatomic bond length v.s. log errors are shown on the energy levels of ground, triplet, singlet, and doubly excited states on hydrogen molecule, respectively. Local search method is (A)Powell, (B)BFGS, (C)Nelder-Mead and (D)Newton's methods. }\label{genera}
\end{figure*}

Secondary, we show the result of the calculation by GMIG-VQE method in Fig. \ref{gene}. Each state of hydrogen molecules for diatomic bond length from 0.1$(\AA)$ to 2.5$(\AA)$ in 0.1 pitch is calculated by GMIG-VQE using (A)Powell, (B)BFGS, (C)Nelder-Mead and (D)Newton's methods as LS. The number of iterations of each method is 500 for Powell methods, 22 for the BFGS method, and 1000 for Newton and Nelder-Mead method, respectively. Log error as a function of r are shown in Fig. \ref{genera} on each methods. Using original data shown in Fig. \ref{genera}, the average log error was shown in Table. \ref{tge}.

\begin{figure*}[h]
\includegraphics[scale=0.45]{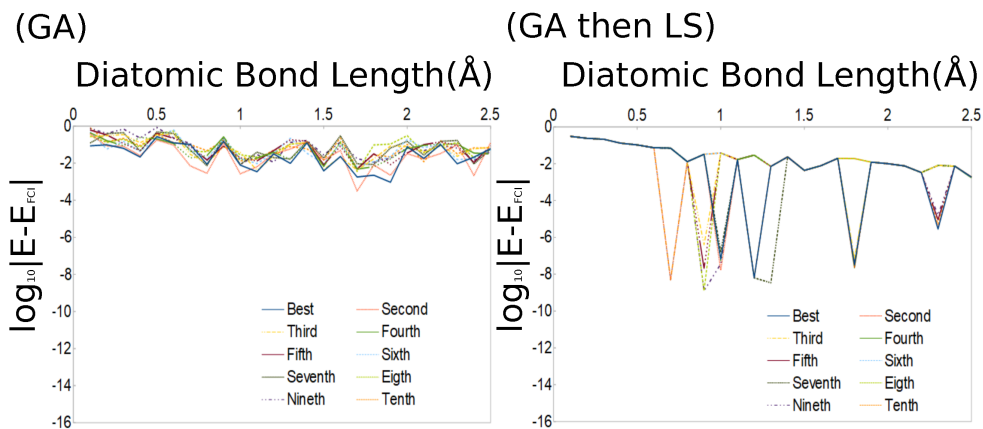}
\newline

\caption{The diatomic bond length v.s. log error of the ten candidates on the energy levels of doubly excited states on hydrogen molecule when GA ends and GA then LS, respectively. The LS is Newton's method.}\label{Na}
\end{figure*}

All cases calculated by GMIG-VQE match the exact values closer than that of VQE with VQD. (C)Nelder-Mead has the largest log error for all states except doubly excited states and (D)Newton has the smallest log error for the doubly excited state. Besides, the accuracy of this state is beyond the chemical accuracy on average. This is because there are 7 points whose log error is below -6. It indicates that GA prepares the close initial values to the global minimum. In addition, the accuracy of the calculation result of (B)BFGS has the second smallest log error for all states except the singlet state. GA then LS improves the accuracy of VQE even the magnitude of it depends on the type of LS. We show the log errors of the calculation result of 10 individuals on the doubly excited state for diatomic bond lengths in case LS is Newton's method when GA ends and when entire GA then LS ends, respectively in Fig. \ref{Na}. Log errors on $r\in[0.1,0.5]$ when GA ends are less than that when entire GA then LS ends. Besides, the individuals that have the smallest log error are seldom the first individual. Rather, tenth individuals sometimes have the smallest log error. Hence, it is confirmed that GA can prepare the initial values close to the global minimum. According to the log errors of when GA then LS ends, four in ten individuals reach the global minimums sometimes. It is supposed that the number of candidates is large enough to reach the global minimums.
It is because the single iteration of BFGS takes more than 10 times larger than the Newton's method. Total time for calculation are shown in Table. \ref{times}. The calculation time is more than 3 hours. Newton is about 3 hours and faster than Powell.
The comparison of GA and GA then LS is shown in Fig. \ref{Na}. Here, Newton is used as LS. As a whole, it suggest that GA then LS is superior to GA.

\begin{table}[h]
\caption{
The average log error of ground, triplet, singlet, and doubly excited state on hydrogen molecule for diatomic bond length r calculated by GMIG-VQE in case (A), (B), (C), and (D), respectively.
}\label{tge}
\begin{tabular}{c|c|c|c|c} \hline \hline
method&Ground&Triplet&Singlet&Doubly \\ \hline
Powell&-8.125&-3.4672&-1.9508&-1.7469 \\
BFGS&-11.0148&-9.7884&-2.5582&-1.9223 \\
Nelder-Mead&-13.4374&-3.7673&-1.8387&-2.0008 \\
Newton&-10.008&-9.9543&-2.1402&-3.3772 \\ \hline
\end{tabular}
\end{table}

\begin{table}[h]\caption{
Total time for calculation in case (A), (B), (C), and (D), respectively. The unit of time is second.
}\label{times}

\begin{tabular}{c|c|c|c|c} \hline \hline

method&Powell&BFGS&Nelder-Mead&Newton \\ \hline
Total time.&14153.204&14237.898&11650.9539&12172.8072 \\ \hline

\end{tabular}
\end{table}

\section{Concluding remarks}\label{4}
In this work, Newton and BFGS method denote higher accuracy conpared to others as methods on LS of GMIG-VQE. These methods use the Hessian and quasi-Hessian matrix for optimization. Hence, second derivative is crucial for optimization on VQE and GA is able to prepare the initial values of LS that is close to its grobal minimums.
BFGS method is a kind of Newton method. Therefore, Newton's method as LS could calculate the energy levels more accurately than normal VQE method. The first issue is how to improve the time for the calculation of the GA method. The bulk of the total calculation time is of the GA method. Besides, the effect of noises must be investigated for future use in real quantum computers. The second issue is the simulation of calculation taking noises into account. We need the comparison with other methods such as SSVQE and MCVQE.
\newpage\bibliographystyle{apsrev4-2}
\bibliography{temp6}

\end{document}